\newif\ifshortver
\shortverfalse

\ifshortver
\documentclass[conference,letterpaper]{IEEEtran}

\else
\documentclass[english,onecolumn]{IEEEtran}
\fi

\usepackage[utf8]{inputenc} 
\usepackage[T1]{fontenc}
\usepackage{url}
\usepackage{ifthen}
\usepackage{cite}
\usepackage[cmex10]{amsmath} 
\usepackage{graphicx}
\usepackage{placeins}
\usepackage{float}
\usepackage{subfigure}
\usepackage{pgfplots}
\usepackage{pgfplotstable}	
\usepackage{amssymb}
\usepackage{color}

\newtheorem{thm}{\protect\theoremname}
\providecommand{\theoremname}{Theorem}

\newtheorem{definition}{\protect\definitionname}
\providecommand{\definitionname}{Definition}

\newtheorem{lemma}{\protect\lemmaname}
\providecommand{\lemmaname}{Lemma}

\begin{document}
\title{Nonasymptotic Oblivious Relaying and Variable-Length Noisy Lossy Source Coding}



\IEEEoverridecommandlockouts

\ifshortver
\else
\author{%
  \IEEEauthorblockN{Yanxiao Liu,  \textit{Member, IEEE}, Sepehr Heidari Advary, and Cheuk Ting Li, \textit{Member, IEEE}\\}
\thanks{\textcolor{black}{The work of Yanxiao Liu was performed when he was with the Department of Information Engineering at The Chinese University of Hong Kong, Hong Kong, China. He is now with the Information Processing and Communications Lab at Imperial College London, London, UK.}
Cheuk Ting Li is with the Department of Information Engineering, The Chinese University of Hong Kong, Hong Kong, China. }
\thanks{This paper was presented in part at the 2025 IEEE International Symposium on Information Theory (ISIT).}
\thanks{
This work was partially supported by two grants from the Research Grants Council of the Hong Kong Special Administrative Region, China [Project No.s: CUHK 24205621 (ECS), CUHK 14209823 (GRF)]. 
}
}
\fi

\maketitle

\begin{abstract}

The information bottleneck channel (or the oblivious relay channel) concerns a channel coding setting where the decoder does not directly observe the channel output. Rather, the channel output is relayed to the decoder by an oblivious relay (which does not know the codebook) via a rate-limited link. The capacity is known to be given by the information bottleneck. We study finite-blocklength achievability results of the channel, where the relay communicates to the decoder via fixed-length or variable-length codes. These two cases give rise to two different second-order versions of the information bottleneck. Our proofs utilize the nonasymptotic noisy lossy source coding results by Kostina and Verd\'{u}, the strong functional representation lemma, and the Poisson matching lemma. Moreover, we also give a novel nonasymptotic variable-length noisy lossy source coding result. 

\end{abstract}

\begin{IEEEkeywords}
Finite blocklength, oblivious relay channel, lossy source coding, channel simulation, network information theory.  
\end{IEEEkeywords}

\section{Introduction}

\begin{figure}
    \centering 
    \includegraphics[scale = 0.95]{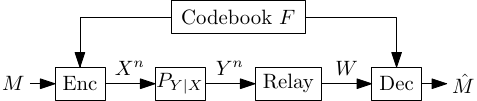} 
    \caption{Information bottleneck channel (or oblivious relaying).}
    \label{fig:info_btl} 
\end{figure}

In the \emph{oblivious relay channel}~\cite{sanderovich2008communication, simeone2011codebook, aguerri2019capacity} (also referred to as the \emph{information bottleneck channel}~\cite{dikshtein2023mismatched, wu2024achievable}), the encoder encodes a message $M$ into a sequence $X^{n}=(X_{1},\ldots,X_{n})$ via random coding and sends it through a memoryless channel $P_{Y|X}$. An oblivious relay receives $Y^{n}$ and transmits a description $W$ to the decoder, which attempts to decode $M$. 
See Figure~\ref{fig:info_btl}
for an illustration. The relay is \emph{oblivious} in the sense that it does not know the random codebook used by the encoder and the decoder, \textcolor{black}{though it knows the distribution of the random codebook, which is assumed to be generated in an i.i.d. manner following an input distribution $P_X$}. As shown by~\cite{sanderovich2008communication}, the minimum asymptotic description rate (as the blocklength $n\to\infty$) needed to support a message transmission rate $\mathsf{C}$ is given by the \emph{information bottleneck} \cite{tishby1999information}, also known as the \emph{relevance-compression function} \cite{slonim2002information}
\begin{equation}
\mathrm{IB}_{X\to Y}(\mathsf{C}):=\min_{P_{U|Y}:\,I(X;U)\ge\mathsf{C}}I(Y;U),\label{eq:ib_max}
\end{equation}
where we assume $X\to Y\to U$ forms a Markov chain,\footnote{\textcolor{black}{For a cardinality bound on the auxiliary random variable $U$, it suffices to consider $|\mathcal{U}|\leq |\mathcal{Y}|+1$ \cite{hsu2018generalizing}.}} \textcolor{black}{and the joint distribution of $X,Y$ is induced by the input distribution $P_X$ and the channel $P_{Y|X}$.}

In this paper, which is the complete version of~\cite{info_btl_isit},\footnote{Compared to the conference paper~\cite{info_btl_isit}, this full version includes the complete proofs of all the theorems, as well as updated descriptions in the literature review and the main sections. New figures have also been provided.}
to study the trade-off between description rate and message rate when the blocklength $n$ is limited, we show a second-order achievability result for the information bottleneck channel in terms of a natural second-order version of the information bottleneck, which we call the \emph{var-information bottleneck}:
\begin{equation}
\mathrm{VIB}_{X\to Y}(\mathsf{C}):=\mathrm{Var}\big[\iota_{Y;U}(Y;U)-\lambda^{*}\iota_{X;U}(X;U)\big],\label{eq:VIB}
\end{equation}
where $\iota_{Y;U}(y;u)=\log\frac{P_{U|Y}(u|y)}{P_{U}(u)}$ is computed by the optimal $P_{U|Y}$ in \eqref{eq:ib_max}, $\iota_{X;U}$ is similar, and $\lambda^{*}:=\frac{\mathrm{d}}{\mathrm{d}\mathsf{C}}\mathrm{IB}_{X\to Y}(\mathsf{C})$.\footnote{\textcolor{black}{If there are multiple $P_{U|Y}$'s attaining the minimum in \eqref{eq:ib_max}, choose the one minimizing the variance in \eqref{eq:VIB}. It suffices to consider $|\mathcal{U}| \leq |\mathcal{Y}| + 2$, which will be shown in Appendix~\ref{subsec:pf_card}.}}

For fixed-length description, we show that a rate 
\begin{equation}
\mathrm{IB}(\mathsf{C}) +\sqrt{\frac{1}{n}\mathrm{VIB}(\mathsf{C})}Q^{-1}(\epsilon)+O\left(\frac{\log n}{n}\right)\label{eq:fl_finite}
\end{equation}
suffices when the blocklength is $n$ and the error probability is $\epsilon$, where $Q^{-1}(\cdot)$ is the inverse of the $Q$-function. This is shown by using the Poisson matching lemma \cite{li2021unified}, and by relating the information bottleneck channel to noisy lossy source coding, where we can utilize the second-order results in \cite{kostina2016nonasymptotic}.

We also study a setting where the description sent by the relay can be variable-length and encoded by a prefix-free code. In this case, the second-order achievability result is instead given in terms of the \emph{conditional-var-information bottleneck}:\footnote{\textcolor{black}{Again, we consider $P_{U|Y}$ attaining the minimum in \eqref{eq:ib_max}, and choose the one minimizing \eqref{eq:CVIB} in case of ties. It suffices to consider $|\mathcal{U}| \leq |\mathcal{Y}| + 2$, which will be shown in Appendix~\ref{subsec:pf_card}.}}
\begin{equation}
\mathrm{CVIB}_{X\to Y}(\mathsf{C}):=\mathbb{E}\big[\mathrm{Var}\big[\lambda^{*}\iota_{X;U}(X;U)\,\big|\,Y,U\big]\big].\label{eq:CVIB}
\end{equation}
Note that CVIB is generally smaller than VIB since VIB is the variance of $\iota(Y;U)-\lambda^{*}\iota(X;U)$, whereas CVIB is its (expected) conditional variance given $Y,U$. We show that for variable-length description, it suffices to use a description rate
\begin{equation}
(1-\epsilon)\bigg(\mathrm{IB}(\mathsf{C})+\sqrt{\frac{\ln n}{n}\mathrm{CVIB}(\mathsf{C})}\bigg)+O\left(\frac{1}{\sqrt{n}}\right). \label{eq:vl_finite}
\end{equation}
Comparing \eqref{eq:fl_finite} and \eqref{eq:vl_finite}, we see that variable-length and fixed-length have vastly different finite blocklength behavior.

These results are proved using techniques in noisy lossy source coding~\cite{dobrushin1962information}, where we have a 2-discrete memoryless source $X^n,Y^n$, the encoder observes $Y^n$ and sends a description to the decoder, which recovers $Z^n$ and aims at having a small probability of excess distortion $\mathbb{P}(d(X^n,Z^n)>\mathsf{D})\le \epsilon$. 
Given by~\cite{dobrushin1962information}, the optimal asymptotic description rate is $R(\mathsf{D}):=\min_{P_{Z|Y}:\,\mathbb{E}[d(X,Z)]\le\mathsf{D}}I(Y;Z)$.
A second-order characterization for the optimal \textcolor{black}{rate} of the fixed-length description
\textcolor{black}{
\begin{equation}
    R(\mathsf{D}) + \sqrt{\frac{1}{n}\tilde{\mathrm{V}}(\mathsf{D})}Q^{-1}(\epsilon)+O\Big(\frac{\log n}{n}\Big),
\end{equation}
}%
was shown in \cite{kostina2016nonasymptotic}, where $\tilde{\mathrm{V}}(\mathsf{D}):=\mathrm{Var}[\iota_{Y;Z^{*}}(Y;Z^{*})+\lambda^{*}d(X,Z^{*})]$, $P_{Z^{*}|Y}$ attains the minimum in $R(\mathsf{D})$, and $\lambda^{*}:=-R'(\mathsf{D})$.
In this paper, we show that for a variable-length description, we can achieve 
\textcolor{black}{a rate (expected length divided by $n$)}
\textcolor{black}{
\begin{equation}
    (1-\epsilon)\bigg(R(\mathsf{D})+\sqrt{\frac{\ln n}{n}\widetilde{\mathrm{CV}}(\mathsf{D})}\bigg)+O\Big(\frac{1}{\sqrt{n}}\Big),
\end{equation}
}%
where $\widetilde{\mathrm{CV}}(\mathsf{D}):=(\lambda^{*})^{2}\mathbb{E}[\mathrm{Var}[d(X,Z^{*})\,|\,Y,Z^{*}]]$. 
This is proved by techniques in \cite{kostina2016nonasymptotic} and Poisson functional representation \cite{sfrl_trans,li2021unified}.

\section{Related Works}

\subsection{Information Bottleneck Channel and Oblivious Relaying}

The setting of oblivious relay processing~\cite{sanderovich2008communication, simeone2011codebook, aguerri2019capacity} was motivated by the architecture of modern communication networks (cloud radio access networks~\cite{peng2015fronthaul}), where access points are connected to a central server via rate-limited, error-free links. 
The scenario involving multiple oblivious relays was investigated  by~\cite{sanderovich2008communication} (also see a $2$-cascaded-relays setting~\cite{ensan2021cloud}), while~\cite{simeone2011codebook} explored cases where encoders can switch among different codebooks and relay nodes have access to certain scheduling information. 
Driven by the lack of knowledge about codebooks, mismatched decoding at the decoder and mismatched compression at the relay were examined in~\cite{dikshtein2023mismatched}. 
For the information bottleneck channel, error exponents have been studied in~\cite{wu2024achievable} and~\cite{wu2026error}, where the former considers the channel under constant-composition codes through random generation of compress-forward codebooks, while the latter utilizes the type covering lemma~\cite{csiszar2011information} with refined analyses using the method of types.
Furthermore, the results of oblivious relaying are closely tied to the information bottleneck problem~\cite{tishby2000information}, which has been extensively studied over the past two decades due to its strong connections to machine learning~\cite{zaidi2020information, goldfeld2020information}. 
The solution to the information bottleneck problem~\eqref{eq:ib_max} aligns with the capacity of the oblivious relay channel and also with the noisy lossy source coding problem~\cite{dobrushin1962information, witsenhausen1980indirect} under a logarithmic distortion function~\cite{courtade2013multiterminal}.

\subsection{Noiseless and Noisy Lossy Source Coding}

In the conventional noiseless lossy source coding setting, the encoder observes the source directly (i.e., $X^n = Y^n$ in the noisy lossy source coding setting). The optimal length for a fixed-length code was characterized to the second-order in \cite{ingber2011dispersion, kostina2012lossy}, based on the rate-dispersion function \cite{kontoyiannis2000pointwise}. For variable-length codes, the study of $d$-semifaithful codes concerns the setting where the distortion is bounded by $\mathsf{D}$ almost surely \cite{ornstein1990universal,yu1993rate,zhang1997redundancy,kontoyiannis2000pointwise,kontoyiannis2002arbitrary}. 
\ifshortver
\else
The optimal expected length is $nR(\mathsf{D})+O(\log n)$ \cite{yu1993rate,zhang1997redundancy}.
Pointwise bounds on the length have been studied in \cite{kontoyiannis2000pointwise}, where the length is shown to be lower-bounded by $nR(\mathsf{D}) + \sqrt{n}G_n + O(\log n)$ where $G_n$ approaches a Gaussian random variable.
\fi
\ifshortver
\else
Variable-length codes allowing a positive probability of excess distortion $\epsilon$ have been studied in \cite{koga2005asymptotic,kostina2015variable}. 
It was shown in \cite{kostina2015variable} that the optimal expected length approaches $n(1-\epsilon)R(\mathsf{D})$ from below.
\fi

The noisy lossy source coding problem (where the encoder only has a noisy observation $Y^n$ of the source $X^n$), also referred to as remote lossy source coding, was first introduced 
\ifshortver
\else
by Dobrushin and Tsybakov~\cite{dobrushin1962information}. 
\fi
It was shown to be asymptotically equivalent to the rate-distortion problem with a surrogate distortion between the output of the noisy channel and the decoder's output, an idea that was further explored in~\cite{witsenhausen1980indirect}. 
It reduces to the information bottleneck problem~\cite{tishby1999information} under the logarithmic distortion measure~\cite{courtade2013multiterminal}. 
\ifshortver
\else
For studies on the strong converse exponent, see~\cite{wu2025strong} and references therein. 
In~\cite{stavrou2023indirect}, the problem was extended to incorporate an $f$-separable distortion measure. 
Refer to~\cite{kittichokechai2016privacy} for further extensions with privacy constraints.
\fi

Finite-blocklength achievability and converse bounds have been investigated in~\cite{kostina2016nonasymptotic}, where it was demonstrated that the dispersion function from~\cite{kostina2012fixed} can be adapted to obtain nonasymptotic results for the noisy lossy source coding setting. 
A converse bound for variable-length noisy lossy source coding was derived in~\cite{saito2018new}. 
For additional nonasymptotic studies on this problem, 
\textcolor{black}{see~\cite{yang2024indirect, zhou2023finite,li2025new}} and references therein.

\ifshortver
\else
\subsection{Channel Simulation and Poisson Functional Representation}
\fi

Channel simulation aims to determine the minimal communication required over a noiseless channel to simulate a given channel $P_{Y|X}$. 
Various settings of channel simulation have been explored~\cite{bennett2002entanglement, bennett2014reverse, cuff2013distributed}. 
One-shot exact channel simulation with unlimited common randomness 
\ifshortver
\else
has been shown to require an average communication cost  $I(X; Y) + O(\log I(X; Y) )$,  by using 
\fi
rejection sampling~\cite{harsha2010communication, flamich2023greedy}, Poisson functional representation~\cite{sfrl_trans}, or a combination of both~\cite{theis2022algorithms}. 
\ifshortver
\else
Together with other related techniques, the Poisson functional representation~\cite{sfrl_trans} has been applied to various settings, including network information theory~\cite{li2021unified, liu2025one, liu2024hiding}, differential privacy~\cite{liu2024universal}, minimax learning~\cite{li2020minimax} and neural compression~\cite{lei2022neural}. 
\fi
\ifshortver
\else
Readers are referred to~\cite{li2024channel} for a comprehensive review on the channel simulation problem. 
\fi 
Moreover, for the information bottleneck channel, a scheme similar to channel simulation has been employed in~\cite{wu2024achievable} to derive the achievable error exponents. By utilizing random coding and uniform index selection, the authors of~\cite{wu2024achievable} demonstrated that the relay's compression scheme can be asymptotically viewed as a discrete memoryless channel $P_{U|Y}$.

\paragraph*{Notations}

All random variables are discrete with finite support unless otherwise stated. 
Entropy is in bits and $\log$ is to the base $2$. 
Write $[n]:=\{1,\ldots,n\}$, $X^{n}=(X_{1},\ldots, X_{n})$. 
Write $D(P\Vert Q)$ for the relative entropy, and $\iota_{X;Y}(x;y)=\log\frac{P_{X,Y}(x,y)}{P_{X}(x)P_{Y}(y)}$ for the information density. 
Write $\{0,1\}^{*} := \cup_{k=0}^\infty \{0,1\}^{k}$ for the set of bit sequences with any length. 
For a set $\mathcal{X}$, $|\mathcal{X}|$ denotes the cardinality. For a sequence $W \in \{0,1\}^{*}$, $|W|$ denotes its length. 
$Q(t):=\mathbb{P}(Z\ge t)$ where $Z\sim N(0,1)$, and $Q^{-1}(\epsilon)$ is its inverse function. 
``Almost surely'' is often abbreviated as ``a.s.''.

{\color{black}
\section{Preliminaries\label{sec:prelim}}
\label{sec:pre}

In this section, we review the Poisson functional representation~\cite{sfrl_trans,li2021unified}, the Poisson matching lemma~\cite{li2021unified} and the corresponding channel simulation result, which will be used extensively throughout this paper.

\medskip
\begin{definition}[Poisson functional representation (PFR)~\cite{sfrl_trans,li2021unified}]\label{def::PFR}
Fix a reference distribution $P_{\bar{X}}$ over $\mathcal{X}$.
Let $0 < T_1 < T_2,\ldots$ be a Poisson process with rate $1$ (i.e., $T_1, T_2-T_1, T_3-T_2,\ldots \stackrel{\mathrm{iid}}{\sim} \mathrm{Exp}(1)$), which is independent of $\bar{X}_1, \bar{X}_2,\ldots \stackrel{\mathrm{iid}}{\sim} P_{\bar{X}}$. 
For any input distribution $Q$ over $\mathcal{X}$ that is absolutely continuous with respect to $P_{\bar{X}}$, the Poisson functional representation (PFR)~\cite{sfrl_trans} selects the point $\tilde{X}_Q := \bar{X}_{K_Q}$ where\footnote{\textcolor{black}{Ties in $\mathrm{argmin}_i \tilde{T}_i$ happen with probability $0$, and can be ignored. We consider discrete distributions throughout this paper. For general distributions, we consider the Radon-Nikodym derivative instead of $Q(\bar{X}_i)/P_{\bar{X}}(\bar{X}_i)$ \cite{sfrl_trans,li2021unified}.}} $K_Q := \mathrm{argmin}_i \tilde{T}_i$,
\begin{equation}
    \tilde{T}_i := \frac{T_i}{Q(\bar{X}_i)/P_{\bar{X}}(\bar{X}_i)}.\label{eq::PFR_K}
\end{equation}
More generally, the generalized Poisson functional representation (GPFR)~\cite{li2021unified} produces a sequence $\tilde{X}_Q(j) := \bar{X}_{K_Q(j)}$ for $j \in \mathbb{N}$, where $K_Q(j)$ is the index of the $j$-th smallest $\tilde{T}_i$, i.e., $|\{i:\tilde{T}_i \leq \tilde{T}_{K_Q(j)}\}| = j$.\footnote{\textcolor{black}{With probability $1$, the sequence $\tilde{T}_i$ has distinct entries, so we can ignore the event of ties.}}
\end{definition}

\medskip

The PFR is related to the A* sampling algorithm \cite{maddison2014sampling,maddison2016poisson} (see \cite{flamich2022fast}).
Given a reference distribution $P_{\bar{X}}$, the PFR draws a random sequence $(\bar{X}_i)_i$ from $P_{\bar{X}}$ and a sequence of times $(T_i)_i$ from a Poisson process. 
These sequences support a ``query operation'' where, upon inputting a distribution $Q$, produces a sample $\tilde{X}_Q$.
If one selects the point $\bar{X}_i$ with the smallest $T_i$, the resulting sample follows $P_{\bar{X}}$. To instead generate a sample from $Q$, the PFR rescales each time as $\tilde{T}_i = T_i /(Q(\bar{X}_i)/P_{\bar{X}}(\bar{X}_i))$, and then selects the point with the smallest $\tilde{T}_i$~\cite{sfrl_trans}. 
More generally, the sequence produced by GPFR is i.i.d. following $Q$ \cite{li2021unified}.

\medskip
\begin{lemma}[\cite{li2021unified}]\label{lem:pfr_dist}
The outputs of the GPFR have a distribution $\tilde{X}_Q(1),\tilde{X}_Q(2),\ldots \stackrel{\mathrm{iid}}{\sim} Q$.
\end{lemma}

\medskip

In communication settings (e.g., \cite{li2021unified, liu2025one, khisti2024unequal, liu2024hiding}), the query operation can be used as the encoding and/or decoding functions. For example, a simple one-shot channel coding scheme for sending a message $M \sim \mathrm{Unif}([\mathsf{L}])$ through a channel $P_{Y|X}$ with input distribution $P_X$ described in~\cite{li2021unified} is to use the PFR with reference distribution $P_X$. 
Suppose the encoder and the decoder share a random codebook $F = (\bar{X}_m)_{m \in [\mathsf{L}]}$ with entries i.i.d. generated from $P_X$. To turn this into the process in PFR, we generate $\bar{X}_m$ i.i.d. from $P_X$ for $m > \mathsf{L}$, and generate a Poisson process $(T_i)_i$, which are assumed to be available to the encoder and the decoder.
The encoding function is $X= \tilde{X}_{P_X}(M)$ (using GPFR), and the decoding function is $\hat{M} = K_Q$ (using PFR) where $Q(x) = P_{X|Y}(x|Y) \propto P_X(x) P_{Y|X}(Y|x)$. The error probability of this scheme can be bounded by the following result called the generalized Poisson matching lemma \cite{li2021unified}.

\medskip
\begin{lemma}[Generalized Poisson matching lemma \cite{li2021unified}]\label{lem:general_pml}
For $m \in \mathbb{N}$, distributions $P,Q$ absolutely continuous with respect to $P_{\bar{X}}$, and $x \in \mathcal{X}$ with $P(x),Q(x)>0$, we have
\begin{align}
& \mathbb{P}\big( K_Q \neq K_P(m) \,\big|\, \tilde{X}_{P}(m)=x \big) \nonumber \\
& \leq 1 - \Big(1-\min \Big\{ \frac{P(x)}{Q(x)},\,1 \Big\}\Big)^m \nonumber \\
& \leq  m \frac{P(x)}{Q(x)}.
\end{align}
\end{lemma}
\medskip
Using this bound, it has been shown in \cite[Theorem 1]{li2021unified} that there is a one-shot channel coding scheme for sending the message $M \sim \mathrm{Unif}([\mathsf{L}])$ through a channel $P_{Y|X}$ with input distribution $P_X$, with average error probability
\begin{align}
&\mathbb{P}(M\neq\hat{M})\le\mathbb{E}\big[1-(1-\min\{2^{-\iota_{X;Y}(X;Y)},1\})^{(\mathsf{L}+1)/2}\big]. \label{eq:one_channel_coding}
\end{align}
This follows from Lemma~\ref{lem:general_pml} and Jensen's inequality since $P_X(x)/P_{X|Y}(x|y) = 2^{-\iota_{X;Y}(x;y)}$. If we only allow the encoder and the decoder to share the random codebook $F = (\bar{X}_m)_{m \in [\mathsf{L}]}$, then we have to fix the values of $\bar{X}_m$ for $m > \mathsf{L}$ and $(T_i)_i$, which can be performed without increasing the probability of error in \eqref{eq:one_channel_coding}.\footnote{\textcolor{black}{There exist fixed values $(\bar{x}_m)_{m> \mathsf{L}}$ and $(t_i)_i$ such that $\mathbb{P}(M\neq\hat{M}) = \mathbb{E}[\mathbb{P}(M\neq\hat{M} | (\bar{X}_m)_{m> \mathsf{L}}, (T_i)_i)] \geq \mathbb{P}(M\neq\hat{M} | (\bar{X}_m)_{m> \mathsf{L}} = (\bar{x}_m)_{m> \mathsf{L}}, (T_i)_i=(t_i)_i)$.}}
This bound is similar to the dependence testing bound \cite{polyanskiy2010channel} (also see \cite{hayashi2009information}).

Moreover, the PFR can be used as a one-shot channel simulation scheme~\cite{sfrl_trans}, where an encoder and a decoder simulate a channel $P_{X|Y}$ with input distribution $P_Y$ in a distributed manner~\cite{harsha2010communication,bennett2002entanglement,cuff2013distributed}. Assume that the encoder and decoder share the common randomness $(\bar{X}_i)_i, (T_i)_i$, generated by the PFR with reference distribution $P_X$. The encoder observes $Y \sim P_Y$, computes $K=K_{P_{X|Y}(\cdot|Y)} \in \mathbb{N}$ using PFR with $Q(x)=P_{X|Y}(x|Y)$, and sends $K$ to the decoder. The decoder observes $K$ and outputs $X = \bar{X}_{K}$. This way, we can ensure that $X$ follows the conditional distribution $P_{X|Y}$ given $Y$, and hence the channel $P_{X|Y}$ is simulated via the noiseless communication of $K$. The amount of communication is approximately $\mathbb{E}[\log K]$, which can be bounded by the following result in \cite{li2024lossy_arxiv} and \cite[Lemma 12]{li2024channel}, which strengthens \cite{sfrl_trans}.

\medskip
\begin{lemma}[\cite{li2024lossy_arxiv,li2024channel}]\label{lem:pfr_logK}
For the index $K_Q$ produced by PFR,
\begin{align}
& \mathbb{E}[\log K_Q] \leq D(Q \Vert P_{\bar{X}}) + 1.
\end{align}
\end{lemma}
\medskip

Using this bound, it has been shown in \cite{li2024lossy_arxiv,li2024channel} that $K_{P_{X|Y}(\cdot|Y)}$ can be encoded using a prefix code with expected length at most $I(Y;X)+ \log(I(Y;X)+2)+3$. This result is also called the strong functional representation lemma \cite{sfrl_trans}. Also see \cite{li2025discrete} for a stronger bound.

}

\section{Noisy Lossy Source Coding}

\begin{figure}[htpb]
    \centering 
    \includegraphics[scale = 0.23]{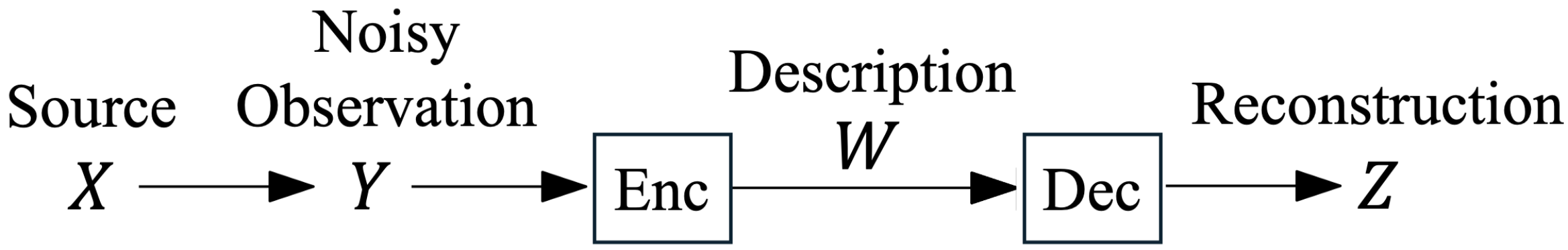} 
    \caption{One-shot noisy lossy source coding.}
    \label{fig:info_nlsc} 
\end{figure}

We first study noisy lossy source coding \cite{dobrushin1962information}, which will be utilized in our analyses on the information bottleneck channel. 
In the one-shot noisy lossy source coding setting, we have a pair of random variables $(X,Y)\sim P_{X,Y}$, where $X\in\mathcal{X}$ is the source, and $Y\in\mathcal{Y}$ is the observation, \textcolor{black}{and we assume that $\mathcal{X}$ and $\mathcal{Y}$ are finite}. 
The encoder observes $Y$ and produces a description $W=f(S,Y)$, where $S\sim P_{S}$ (independent of $X,Y$) is the encoder's local randomness.\footnote{The local randomness $S$ is not useful for fixed-length settings, but can be useful in variable-length cases to randomize between two encoding functions.} The decoder observes $W$ and recovers $Z=g(W)\in\mathcal{Z}$. The goal is to have a small probability of excess distortion $P_{e}:=\mathbb{P}(d(X,Z)>\mathsf{D})$, where $d: \mathcal{X}\times\mathcal{Z} \to\mathbb{R}$ is the distortion measure, and $\mathsf{D}\in\mathbb{R}$. The setting is illustrated in Figure~\ref{fig:info_nlsc} as follows. 

For the fixed-length setting where we require $W\in[\textcolor{black}{\mathsf{L}_W}]$, this problem has been studied in \cite{kostina2016nonasymptotic} by Kostina and Verd\'{u}, which gives the following theorem.  
\textcolor{black}{We note that in Theorem~\ref{thm:noisy_fl}, $P_{\bar{Z}}$ is an auxiliary distribution defined on the space of the decoder’s output, and the result holds for any choice of $P_{\bar{Z}}$ on that space. The same remark applies to other parts of the paper where similar arguments are used. }
\begin{thm}
[\cite{kostina2016nonasymptotic}]\label{thm:noisy_fl}For any $P_{\bar{Z}}$ and $\gamma>0$, there exists a fixed-length code with
\begin{equation}
    P_{e}\le\mathbb{P}\big(\psi_{\bar{Z}}(Y,\mathsf{D},T)\ge\log\gamma\big)+e^{-\textcolor{black}{\mathsf{L}_W}/\gamma},
\end{equation}
where $T\sim\mathrm{Unif}(0,1)$ is independent of $Y$, and
\begin{equation}
\psi_{\bar{Z}}(y,\mathsf{D},t):=\inf_{P_{Z}:\,\mathbb{P}(d(X,Z)>\mathsf{D}|Z,Y=y)\le t\;\mathrm{a.s.}}D(P_{Z}\Vert P_{\bar{Z}}).\label{eq:psi}
\end{equation}
(We assume $Z{\perp\!\!\!\perp}(X,Y)$ above.)
\end{thm}
\smallskip{}

In this paper, we discuss a variable-length setting where $W\in\mathcal{C}$ lies in a prefix-free codebook $\mathcal{C}\subseteq\{0,1\}^{*}$, and we are interested in minimizing the expected length $\mathbb{E}[|W|]$. We show that the expected length can also be bounded in terms of $\psi_{\bar{Z}}$ in \eqref{eq:psi}, using the technique in \cite{kostina2016nonasymptotic} and the strong functional representation lemma \cite{sfrl_trans,li2024lossy_arxiv}.

\smallskip{}
\begin{thm}
\label{thm:noisy_vl}For any $P_{\bar{Z}}$, $\epsilon'>0$, and function $\beta:\mathcal{Y}\to[0,1]$, there is a variable-length code with $P_{e}\le\mathbb{E}[\beta(Y)]+\epsilon'$ and
\begin{equation}
    \mathbb{E}[|W|]\le\ell\left(\mathbb{E}\big[(1-\beta(Y))\psi_{\bar{Z}}(Y,\mathsf{D},\epsilon')\big]\right), \label{eq:lossy_var_prefix}
\end{equation}
assuming the expectation above is finite,\footnote{In particularly, we must have ``$\psi_{\bar{Z}}(Y,\mathsf{D},\epsilon')<\infty$ or $\beta(Y)=1$'' a.s..} where $\ell(t):=t+\log(t+2)+4$.
\end{thm}
\begin{IEEEproof}
Let $\phi(y,z,\mathsf{D}):=\mathbb{P}(d(X,z)>\mathsf{D}|Y=y)$. 
We use the Poisson functional representation \cite{sfrl_trans,li2021unified} \textcolor{black}{(Definition~\ref{def::PFR})}. 
Let $0<T_{1}<T_{2},\ldots$ be a Poisson process and $\bar{Z}_{1},\bar{Z}_{2},\ldots\stackrel{iid}{\sim}P_{\bar{Z}}$. Consider a channel $P_{\hat{Z}|Y}$, where conditional on $Y=y$, $\hat{Z}$ has the same distribution as $\bar{Z}\sim P_{\bar{Z}}$ conditional on $\phi(y,\bar{Z},\mathsf{D})\le\epsilon'$,
\textcolor{black}{that is, $P_{\hat{Z}|Y}(z|y) = \mathbf{1}\{\phi(y,z,\mathsf{D})\le\epsilon'\} P_{\bar{Z}}(z) / \mathbb{P}(\phi(y,\bar{Z},\mathsf{D})\le\epsilon')$. For values of $y$ where $\mathbb{P}(\phi(y,\bar{Z},\mathsf{D})\le\epsilon')=0$, $P_{\hat{Z}|Y}(z|y)$ is arbitrarily chosen.}
\ifshortver
\else
\textcolor{black}{The Poisson functional representation selects}
\begin{equation}
    K:=\mathrm{argmin}_{k} \frac{T_{k}}{P_{\hat{Z}|Y}(\bar{Z}_{k}|Y) \big/ P_{\bar{Z}}(\bar{Z}_{k})}.
\end{equation}
\fi
\textcolor{black}{By Lemma~\ref{lem:pfr_dist} and Lemma~\ref{lem:pfr_logK} \cite{li2024lossy_arxiv,li2024channel},}
$\bar{Z}_{K}|\{Y=y\}\sim P_{\hat{Z}|Y}(\cdot|y)$, 
\textcolor{black}{and for $y$ with $P_Y(y)>0$ and $\beta(y)<1$ (which implies $\psi_{\bar{Z}}(y,\mathsf{D},\epsilon') < \infty$ and $\mathbb{P}(\phi(y,\bar{Z},\mathsf{D})\le\epsilon')>0$ since the expectation in \eqref{eq:lossy_var_prefix} is assumed to be finite), we have}
\ifshortver
\else
\begin{align}
& \mathbb{E}[\log K|Y=y] \nonumber\\
& \le D(P_{\hat{Z}|Y}(\cdot|y)\Vert P_{\bar{Z}})+1 \nonumber\\
 & = -\log\mathbb{P}(\phi(y,\bar{Z},\mathsf{D})\le\epsilon')+1   \nonumber\\
 &\le  \psi_{\bar{Z}}(y,\mathsf{D},\epsilon')+1, \label{eq:pf_psi_Z}
\end{align}
\fi
where the last inequality is because if $P_{Z}$ satisfies that $\phi(y,Z,\mathsf{D})\le\epsilon$ almost surely, then 
\ifshortver
\else
\begin{align}
 D(P_{Z}\Vert P_{\bar{Z}}) &\ge D(P_{\mathbf{1}\{\phi(y,Z,\mathsf{D})\le\epsilon\}}\Vert P_{\mathbf{1}\{\phi(y,\bar{Z},\mathsf{D})\le\epsilon\}}) \nonumber \\
    & = -\log \mathbb{P}(\phi(y,\bar{Z},\mathsf{D})\le\epsilon)
\end{align}\fi
(this step appeared in \cite{kostina2016nonasymptotic}).
Construct a randomized coding scheme as follows: the encoder observes $Y$, outputs $\tilde{K}=K$ with probability $1-\beta(Y)$, or outputs $\tilde{K}=1$ with probability $\beta(Y)$, and then encodes $\tilde{K}$ using an optimal prefix-free code into $W$. The decoder decodes $\tilde{K}$ from $W$ and outputs $Z=\bar{Z}_{\tilde{K}}$. Since $\bar{Z}_{K}|\{Y=y\}\sim P_{\hat{Z}|Y}(\cdot|y)$, 
\ifshortver
\else
\begin{align}
& \mathbb{P}(d(X,Z)>\mathsf{D}) \; \nonumber\\
& \le \; \mathbb{P}(\tilde{K}\neq K)+\mathbb{P}(d(X,\bar{Z}_{K})>\mathsf{D}) \nonumber\\
 &\le \; \mathbb{E}[\beta(Y)]+\mathbb{E}[\phi(Y,\bar{Z}_{K}, \mathsf{D})] \nonumber \\
 & \le\; \mathbb{E}[\beta(Y)]+\epsilon'.
\end{align}
\fi
We have
\textcolor{black}{
\begin{align}
\mathbb{E}[\log\tilde{K}]  &=\mathbb{E}\left[(1-\beta(Y))\mathbb{E}[\log K|Y]\right] \nonumber \\
 & \le\mathbb{E}\left[(1-\beta(Y))\psi_{\bar{Z}}(Y,\mathsf{D},\epsilon')\right]+1. \label{eq:pf_bound_logK}
\end{align}
}
By the maximum entropy distribution for fixed $\mathbb{E}[\log\tilde{K}]$ \cite{sfrl_trans,li2024channel},\footnote{\textcolor{black}{The maximum entropy distribution for a fixed $\mathbb{E}[\log\tilde{K}]$ is the zeta distribution $P_{\tilde{K}}(k)\propto k^{-\lambda}$ for an appropriate $\lambda>0$ \cite{visser2013zipf,sfrl_trans}. Using this fact, it has been shown in \cite{sfrl_trans} that $H(\tilde{K}) \leq \mathbb{E}[\log\tilde{K}] + \log(\mathbb{E}[\log\tilde{K}] + 1) + 1$.}} 
\ifshortver
\else
we have
\begin{equation}
    H(\tilde{K})\le\ell(\mathbb{E}[(1-\beta(Y))\psi_{\bar{Z}}(y,\mathsf{D},\epsilon')])-2.
\end{equation}
\fi
We can then use Huffman code \cite{huffman1952method} to derive $\mathbb{E}[|W|]\le H(\tilde{K})+1$. 

The remaining problem is that these encoding and decoding functions depend on the common randomness $G:=(\bar{Z}_{i},T_{i})_{i}$. To resolve this, we use the strategy in \cite[Theorem 2]{sfrl_trans} which incurs a $1$-bit penalty on $\mathbb{E}[|W|]$. 
\ifshortver
\else
See Appendix \ref{subsec:pf_noisy_vl}. 
\fi
\end{IEEEproof}
\smallskip{}

\textcolor{black}{We also remark that if the prefix condition for $W$ is removed, i.e., we allow any variable-length codeword $W \in \{0,1\}^*$ which is not necessarily in a prefix-free codebook $\mathcal{C}$, then we can encode $\tilde{K}$ into a codeword of length $\lfloor \log \tilde{K}\rfloor$ \cite{alon1994lower,blundo1996new}. By \eqref{eq:pf_bound_logK}, we can have a non-prefix code with an expected length 
\begin{align}
\mathbb{E}[|W|]  \leq \mathbb{E}\left[(1-\beta(Y))\psi_{\bar{Z}}(Y,\mathsf{D},\epsilon')\right]+1. \label{eq:nonprefix}
\end{align}
This is smaller than the result for prefix codes \eqref{eq:lossy_var_prefix} by a logarithmic term.}

We also study the block setting where $X=X^{n}$, $Y=Y^{n}$, $Z=Z^{n}$ are sequences, $(X_{i},Y_{i})\stackrel{iid}{\sim}P_{X,Y}$, and $d(x^{n},z^{n})=n^{-1}\sum_{i=1}^{n}d(x_{i},z_{i})$. For the fixed-length setting, as $n\to\infty$, the optimal asymptotic description rate is \cite{dobrushin1962information}
\begin{equation}
R(\mathsf{D}):=\min_{P_{Z|Y}:\,\mathbb{E}[d(X,Z)]\le\mathsf{D}}I(Y;Z).
\end{equation}
The following refined bound for the fixed-length setting was shown in \cite[Theorem 5]{kostina2016nonasymptotic}.
\begin{thm}[\cite{kostina2016nonasymptotic}]\label{thm:noisy_fl_2}Under the regularity conditions in \cite{kostina2016nonasymptotic},\footnote{We need $\min\{\mathsf{D}':\,R(\mathsf{D}')<\infty\}<\mathsf{D}<\min_{z}\mathbb{E}[d(X,z)]$, $R(\mathsf{D})$ is twice continuously differentiable as a function of $P_{Y}$ (assuming $P_{X,Y}=P_{X|Y}P_{Y}$), and perturbing $P_{Y}$ within a neighborhood of the original $P_{Y}$ will not affect the support of $Z^{*}$, where $P_{Z^{*}|Y}$ attains the minimum in $R(\mathsf{D})$.} 
\textcolor{black}{letting $\mathsf{L}_W^*(n,\epsilon)$ be the smallest $\mathsf{L}_W$ such that there exists a fixed-length code with blocklength $n$ and $P_{e}\le\epsilon$, we have
\begin{equation}
\frac{\log \mathsf{L}_W^*(n,\epsilon)}{n} = R(\mathsf{D})+\sqrt{\frac{\tilde{\mathrm{V}}(\mathsf{D})}{n}}Q^{-1}(\epsilon)+O\Big(\frac{\log n}{n}\Big),
\end{equation}
}
where $\tilde{\mathrm{V}}(\mathsf{D}):=\mathrm{Var}[\iota_{Y;Z^{*}}(Y;Z^{*})+\lambda^{*}d(X,Z^{*})]$, $P_{Z^{*}|Y}$ attains the minimum in $R(\mathsf{D})$, and $\lambda^{*}:=-R'(\mathsf{D})$.
\end{thm}

\smallskip{}

In this paper, for the variable-length setting, we prove the following bound. 
\ifshortver
\else
The proof is in Appendix \ref{subsec:pf_noisy_vl_2}.
\fi

\smallskip{}
\begin{thm}
\label{thm:noisy_vl_2}Under the regularity conditions in Theorem \ref{thm:noisy_fl_2}, for $\epsilon>0$, if $n\ge n_{0}$ (where $n_{0}$ depends on $P_{X,Y},d,\mathsf{D},\epsilon$), there exists a variable-length code with $P_{e}\le\epsilon$,\footnote{\label{fn:noisy_vl_2_pe}This can be strengthened to $\mathbb{P}(d(X^{n}\!,Z^{n})>\mathsf{D}-n^{-1}\log n)\le\epsilon-1/\sqrt{n}$. \textcolor{black}{See Appendix \ref{subsec:pf_noisy_vl_2} for more details}.} and
\textcolor{black}{
\begin{equation}
    \frac{\mathbb{E}[|W|]}{n} \le(1-\epsilon)\left(R(\mathsf{D})+\sqrt{\frac{\ln n}{n}\widetilde{\mathrm{CV}}(\mathsf{D})}\right)+O\Big(\frac{1}{\sqrt{n}}\Big),
\end{equation}
}
where $\widetilde{\mathrm{CV}}(\mathsf{D}):=(\lambda^{*})^{2}\mathbb{E}[\mathrm{Var}[d(X,Z^{*})\,|\,Y,Z^{*}]]$, $P_{Z^{*}|Y}$ attains the minimum in $R(\mathsf{D})$,\footnote{If there are multiple $P_{Z^{*}|Y}$ attaining the minimum, choose the one that gives the smallest $\mathbb{E}[\mathrm{Var}[d(X,Z^{*})\,|\,Y,Z^{*}]]$.} and $\lambda^{*}:=-R'(\mathsf{D})$. The constant in $O(\sqrt{n})$ can depend on $P_{X,Y},d,\mathsf{D},\epsilon$.
\end{thm}
\smallskip{}

Note that $\widetilde{\mathrm{CV}}(\mathsf{D})\le\tilde{\mathrm{V}}(\mathsf{D})$ since $\tilde{\mathrm{V}}(\mathsf{D})$ is the variance of $\iota(Y;Z^{*})+\lambda^{*}d(X,Z^{*})$, and $\widetilde{\mathrm{CV}}(\mathsf{D})$ is its conditional variance given $Y,Z^{*}$. We observe that the variable-length case in Theorem \ref{thm:noisy_vl_2} exhibits a different behavior compared to the fixed-length case in Theorem \ref{thm:noisy_fl_2}. The asymptotic rate is $(1-\epsilon)R(\mathsf{D})$ instead of $R(\mathsf{D})$, which is similar to the phenomenon observed in \cite{koga2005asymptotic, kostina2015variable} for lossless and lossy source coding with error. Intuitively, we can discard a portion $\epsilon$ of the sequences $Y^{n}$ by assigning the same short codeword to them, which induces an error probability $\epsilon$ while reducing the expected length by $\approx\epsilon R(\mathsf{D})$. Nevertheless, unlike the result for variable-length noiseless (i.e., $X=Y$) lossy source coding in \cite{kostina2015variable} which shows that \textcolor{black}{$\mathbb{E}[|W|]/n=(1-\epsilon)R(\mathsf{D})-\zeta/\sqrt{n}+O((\log n)/n)$} (the constant $\zeta$ is given in terms of $\epsilon$ and the rate-dispersion function) where the rate $\mathbb{E}[|W|]/n$ approaches $(1-\epsilon)R(\mathsf{D})$ from below, the noisy lossy source coding result in Theorem \ref{thm:noisy_vl_2} gives a rate which approaches $(1-\epsilon)R(\mathsf{D})$ from \emph{above}.\footnote{\textcolor{black}{Note that \cite{kostina2015variable} studies variable-length codes without the prefix condition. Since the gap between the optimal expected length of non-prefix codes and that of prefix codes is $O(\log n)$, $\mathbb{E}[|W|]/n=(1-\epsilon)R(\mathsf{D})-\zeta/\sqrt{n}+O((\log n)/n)$ holds regardless of whether the prefix condition is imposed \cite{kostina2015variable}.}} The reason is that we have to take into account of the variance of $d(X^{n},Z^{n})$ which increases with $n$, whereas in the noiseless case $d(X^{n},Z^{n})$ is fixed by $X^{n}=Y^{n}$ and $Z^{n}$. 

\textcolor{black}{If the prefix condition for $W$ is removed as in \eqref{eq:nonprefix}, this will reduce $\mathbb{E}[|W|]$ by at most $O(\log n)$, which will not affect the result in Theorem \ref{thm:noisy_vl_2} since $(\log n)/n = O(1/\sqrt{n})$.}

\section{Information Bottleneck Channel}

In this section, we construct schemes for both the fixed-length and the variable-length cases of the information bottleneck channel, by utilizing results on noisy lossy source coding. 

We first define the one-shot information bottleneck channel. 
An encoder observes the message $M\sim\mathrm{Unif}([\mathsf{L}])$ and a shared random codebook $F\sim P_{F}$,\footnote{$F$ is a random variable that represents a random choice out of the $|\mathcal{X}|^{\mathsf{L}}$ different mappings from $[\mathsf{L}]$ to $\mathcal{X}$.} and sends $X=f(F,M)\in\mathcal{X}$ (where $\mathcal{X}$ is a finite set) through a channel $P_{Y|X}$ which outputs $Y\in\mathcal{Y}$ (where $\mathcal{Y}$ is also finite). 
We require the codebook to be i.i.d., i.e., $f(F,1),\ldots,f(F,\mathsf{L})$ are i.i.d. following an input distribution $P_X$.
An oblivious relay observes $Y$ (but not $F$) and sends a description $W=f_{r}(S,Y)$ noiselessly to the decoder, where $S\sim P_{S}$ is the relay's local randomness. The decoder observes $F$ and $W$, and then recovers $\hat{M}=g(F,W)\in[\mathsf{L}]$. The goal is to minimize the error probability $P_{e}:=\mathbb{P}(M\neq\hat{M})$. 
The setting has been presented in Figure~\ref{fig:info_btl}.

About the description $W$, two settings will be studied: 
\begin{itemize}
    \item Fixed-length description: $W\in[\textcolor{black}{\mathsf{L}_W}]$, and we want to minimize $\textcolor{black}{\mathsf{L}_W}\in\mathbb{N}$; 
    
    \item Variable-length description: $W\in\mathcal{C}_W$ is in a prefix-free codebook $\mathcal{C}_{W}\subseteq\{0,1\}^{*}$, and we want to minimize the expected length $\mathbb{E}[|W|]$. 
\end{itemize} 

Moreover, we also study the block case, where the encoder sends a sequence $X^{n}\in\mathcal{X}$ that is passed through a memoryless channel $P_{Y|X}^{n}$ ($n$ copies of $P_{Y|X}$), so the relay observes a sequence $Y^{n}$, i.e., we substitute $X=X^{n}$, $Y=Y^{n}$ and $P_{Y|X}=P_{Y|X}^{n}$ in the one-shot setting. 
Let $\mathsf{B}_{\mathrm{F}}^{*}(n,\mathsf{C},\epsilon)$ be the smallest possible relay description rate $n^{-1}\log\textcolor{black}{\mathsf{L}_W}$ among fixed-length schemes with message rate $n^{-1}\log\mathsf{L}\ge\mathsf{C}$ and $P_{e}\le\epsilon$. 
\textcolor{black}{
Let $\mathsf{B}_{\mathrm{V}}^{*}(n,\mathsf{C},\epsilon)$ be the smallest possible relay description rate $n^{-1}\mathbb{E}[|W|]$ (i.e., infimum of the set of possible description rates) among variable-length schemes with $n^{-1}\log\mathsf{L}\ge\mathsf{C}$ and $P_{e}\le\epsilon$. 
}

For the asymptotic setting where $n\to\infty$, the capacity for fixed-length description has been characterized in \cite{sanderovich2008communication} as
\begin{equation}
    \mathrm{limsup}_{\epsilon\to 0} \, \mathrm{limsup}_{n\to\infty} \, \mathsf{B}_{\mathrm{F}}^{*}(n,\mathsf{C},\epsilon)=\mathrm{IB}_{X\to Y}(\mathsf{C}).
\end{equation}
\cite{sanderovich2008communication} shows that the same asymptotic limit also holds for $\mathsf{B}_{\mathrm{V}}^{*}$.
In the following subsections, using the techniques in noisy lossy source coding, we present nonasymptotic achievability results for the variable-length and fixed-length settings. Interestingly, the nonasymptotic results for them are rather different.

\subsection{Variable-length Description}

\begin{figure}[htpb]
    \centering 
    \includegraphics[scale = 0.9]{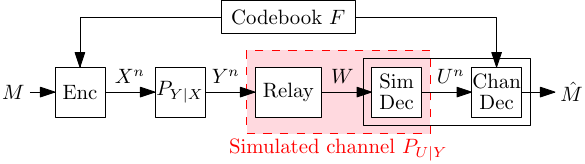} 
    \caption{Coding for the information bottleneck channel via channel simulation.}
    \label{fig:ib_chl_simu} 
\end{figure}

We start with the setting where the relay sends a variable-length description $W$ in a prefix-free codebook. For the one-shot variable-length setting, one straightforward scheme is to consider the $P_{U|Y}$ that achieves the minimum in the information bottleneck \eqref{eq:ib_max}, and have the relay perform a one-shot exact channel simulation scheme (e.g., \cite{harsha2010communication, sfrl_trans,theis2022algorithms, flamich2023greedy}) to simulate $P_{U|Y}$. Channel simulation can be performed with an expected description length bounded by $I(Y;U)+\log(I(Y;U)+2)+3$ \cite{sfrl_trans,li2024lossy_arxiv}. The decoder can then recover $U$, treat $P_{U|X}$ as a usual noisy channel, and perform decoding of the channel code. We illustrate the idea in Figure~\ref{fig:ib_chl_simu}.

The following result is a consequence of the channel simulation result in \cite{sfrl_trans,li2024lossy_arxiv} and the one-shot channel coding result in \cite{li2021unified} (one may also use any of the bounds in \cite{polyanskiy2010channel}).

\smallskip{}
\begin{thm}
\label{thm:ib_one_chansim}Fix any $P_{X}$, $P_{U|Y}$ and $\epsilon'\ge0$. There is a one-shot variable-length scheme with
\begin{equation}
P_{e}\le\mathbb{E}\left[1-\big(1-\min\big\{2^{-\iota_{X;U}(X;U)},1\big\}\big)^{(\mathsf{L}+1)/2}\right]+\epsilon',\label{eq:one_vl_pe}
\end{equation}
and $\mathbb{E}[|W|]\le\ell((1-\epsilon')I(Y;U))$, where $\ell(t):=t+\log(t+2)+4$.
\end{thm}
\begin{IEEEproof}
\textcolor{black}{
Applying the channel simulation construction in Section \ref{sec:prelim} \cite{sfrl_trans,li2024lossy_arxiv} on $P_{U|Y}$,} we have an encoding function $K=f_{a}(S_{a},Y)$ where $K\in\mathbb{N}$ (to be used by the relay) and a decoding function $U=g_{a}(S_{a},K)$ (to be used by the decoder, where $S_{a}$ is \textcolor{black}{the common randomness produced by PFR}) such that $U$ follows the conditional distribution $P_{U|Y}$ given $Y$, and $\mathbb{E}[\log K]\le I(Y;U)+1$ \textcolor{black}{by Lemma~\ref{lem:pfr_logK}}. Let $\tilde{K}=K$ with probability $1-\epsilon'$, and $\tilde{K}=1$ with probability $\epsilon'$. As in the proof of Theorem \ref{thm:noisy_vl}, $\tilde{K}$ can be encoded into $W$ using a prefix-free code with 
\ifshortver
\else
\begin{align}
& \mathbb{E}[|W|] \nonumber\\
& \le \ell(\mathbb{E}[\log\tilde{K}]-1)-1 \nonumber\\
& \le \ell((1-\epsilon')(\mathbb{E}[\log K]-1))-1 \nonumber \\ 
& \le \ell((1-\epsilon')I(Y;U))-1.
\end{align}
\fi
Using the one-shot channel coding result in \cite[Theorem 1]{li2021unified} \textcolor{black}{(see Lemma~\ref{lem:general_pml} and \eqref{eq:one_channel_coding})} on $P_{U|X}$, we have an encoding function $X=f_{b}(F,M)$ (to be used by the encoder) and a decoding function $\hat{M}=g_{b}(F,U)$ (to be used by the decoder, where $F$ is an i.i.d. random codebook) such that $\mathbb{P}(M\neq\hat{M})\le\mathbb{E}[1-(1-\min\{2^{-\iota_{X;U}(X;U)},1\})^{(\mathsf{L}+1)/2}]$.

The scheme for the information bottleneck channel is as follows: the encoder observes $M$ and outputs $X=f_{b}(F,M)$; the relay observes $Y$, computes $K=f_{a}(S_{a},Y)$, generates $\tilde{K}$ and encodes it to $W$; the decoder observes $W$, recovers $\tilde{K}$, $U=g_{a}(S_{a},\tilde{K})$, $\hat{M}=g_{b}(F,U)$. Since $\mathbb{P}(K\neq\tilde{K})\le\epsilon'$, using $\tilde{K}$ instead of $K$ increases $P_{e}$ by at most $\epsilon'$. 

The remaining problem is that there is a common randomness $S_{a}$ shared between the relay and the decoder. It can be removed using the same technique as in the proof of Theorem \ref{thm:noisy_vl}, incurring a $1$-bit penalty on $\mathbb{E}[|W|]$.
\end{IEEEproof}
\smallskip{}

A direct application of Theorem \ref{thm:ib_one_chansim} to the asymptotic setting $X=X^{n}$, $Y=Y^{n}$ yields the asymptotic result $\mathrm{limsup}_{n\to\infty}\mathsf{B}_{\mathrm{V}}^{*}(n,\mathsf{C},\epsilon)\le(1-\epsilon)\mathrm{IB}_{X\to Y}(\mathsf{C})$. 
To refine the bound, we utilize the noisy lossy source coding result in Theorem \ref{thm:noisy_vl}. Intuitively, for any fixed $p_{U|Y}$ (e.g., the one achieving the minimum in $\mathrm{IB}_{X\to Y}(\mathsf{C})=\min_{P_{U|Y}:\,I(X;U)\ge\mathsf{C}}I(Y;U)$), the relay performs a noisy lossy source coding on $Y$ to allow the decoder to recover $\hat{U}$, with a distortion function $d(x,\hat{u})=-\iota_{X;U}(x;\hat{u})$. As long as the distortion is small enough, that is, $\iota_{X;U}(x;\hat{u})\gg\log\mathsf{L}$, the decoder can decode $X$ using $\hat{U}$ via the Poisson matching lemma \cite{li2021unified}.
\textcolor{black}{
Note that $U$ is an auxiliary random variable that is not present in the information bottleneck channel setting, but is introduced as an intermediate step to construct a coding scheme via noisy lossy source coding.}

\smallskip{}
\begin{thm}
\label{thm:ib_one_vl}Fix any $P_{X}$, $P_{U|Y}$, $\mathsf{C},\epsilon'>0$, and function $\beta:\mathcal{Y}\to[0,1]$. There is a one-shot variable-length scheme with message size $\mathsf{L}$,
\ifshortver
\else
\begin{align}
    P_{e} & \le\mathbb{E}[\beta(Y)]+2^{-\mathsf{C}}(\mathsf{L}+1)/2+\epsilon',\\
    \mathbb{E}[|W|] & \le \ell\big(\mathbb{E}\big[(1-\beta(Y))\psi_{U}(Y,\mathsf{C},\epsilon')\big]\big),
\end{align}
\fi
assuming the expectation above is finite, where $\ell(t):=t+\log(t+2)+4$, and
\begin{equation}
\psi_{U}(y,\mathsf{C},t):=\inf_{P_{\tilde{U}}:\,\mathbb{P}(\iota_{X;U}(X;\tilde{U})<\mathsf{C}|\tilde{U},Y=y)\le t\;\mathrm{a.s.}}D(P_{\tilde{U}}\Vert P_{U}).\label{eq:psi_U}
\end{equation}
(We assume $\tilde{U}{\perp\!\!\!\perp}(X,Y)$ above.)
\end{thm}
\begin{IEEEproof}
Fix any $P_{X}$, $P_{U|Y}$. Applying Theorem \ref{thm:noisy_vl} on the distortion function $d(x,\hat{u})=-\iota_{X;U}(x;\hat{u})$ and distortion level $\mathsf{D}=-\mathsf{C}$, we have a code for noisy lossy source coding, with encoding function $W=f_{r}(S,Y)$ (used by the relay) and decoding function $\hat{U}=g_{r}(W)$ (used by the decoder) so that 
\ifshortver
\else
\begin{align}
& \mathbb{P}(d(X;\hat{U})>\mathsf{D}) \nonumber \\
& = \mathbb{P}(\iota_{X;U}(X;\hat{U}) <\mathsf{C}) \nonumber \\ 
& \le \mathbb{E}[\beta(Y)]+\epsilon', \label{eq:pf_ib_dist}
\end{align}
\fi
and $\mathbb{E}[|W|]\le\ell(\mathbb{E}[(1-\beta(Y))\psi_{U}(Y,\mathsf{C}, \epsilon')])$. It is left to design the encoder and the decoder.

We utilize the Poisson functional representation \textcolor{black}{(Definition~\ref{def::PFR})} \cite{sfrl_trans,li2021unified}. Let $\bar{X}_{1},\bar{X}_{2},\ldots\stackrel{iid}{\sim}P_{X}$, and $0<T_{1}<T_{2},\ldots$ be a Poisson process. The encoder observes $M\in[\mathsf{L}]$ and sends $X=\bar{X}_{M}$. The decoder observes $\hat{U}=g_{r}(W)$ and recovers
\ifshortver
\else
\begin{equation}
    \hat{M}:=\mathrm{argmin}_{k\in[\mathsf{L}]}\frac{T_{k}}{P_{X|U}(\bar{X}_{k}|\hat{U})/P_{X}(\bar{X}_{k})}.
\end{equation}
\fi
\textcolor{black}{By the generalized Poisson matching lemma (Lemma~\ref{lem:general_pml}) \cite{li2021unified}},
\ifshortver
\else
\begin{align}
& \mathbb{P}(M\neq\hat{M}|M=m) \nonumber\\
& \textcolor{black}{\leq \,} \mathbb{E}\left[\min\left\{ m\frac{P_{X}(X)}{P_{X|U}(X|\hat{U})},1\right\} \right] \nonumber\\
 &  \textcolor{black}{= \,} \mathbb{E}\left[\min\left\{ m2^{-\iota_{X;U}(X;\hat{U})},1\right\} \right] \nonumber
 \\
 & \le 2^{-\mathsf{C}}m+\mathbb{E}[\beta(Y)]+\epsilon',
\end{align}
\fi
where the last inequality is by \eqref{eq:pf_ib_dist}. The result follows from averaging over $M\sim\mathrm{Unif}([\mathsf{L}])$. We take the codebook random variable to be $F=(\bar{X}_{m})_{m\in[\mathsf{L}]}$.\footnote{$(T_{m})_{m}$ is only a local randomness at the decoder. If this is not allowed, we can fix a particular $(t_{m})_{m}$ that satisfies the bound on $P_{e}$.}
\end{IEEEproof}
\smallskip{}

We then show an achievability result for the block setting in terms of the information bottleneck $\mathrm{IB}(\mathsf{C})$ in \eqref{eq:ib_max} and the conditional var-information bottleneck $\mathrm{CVIB}(\mathsf{C})$ in \eqref{eq:CVIB}, using the noisy lossy source coding result in Theorem \ref{thm:noisy_vl_2}. \textcolor{black}{Refer to Appendix~\ref{subsec:pf_card} for a cardinality bound for the auxiliary random variable $U$ in the definition of $\mathrm{CVIB}(\mathsf{C})$.}

\smallskip{}
\begin{thm}
\label{thm:ib_n_vl}Fix any $P_{X}$, $\epsilon>0$ and $0<\mathsf{C}<I(X;Y)$. 
\textcolor{black}{
Assume that $\tilde{R}(\mathsf{C}):=\min_{P_{\tilde{U}|Y}:\,\mathbb{E}[\iota_{X;U}(X,\tilde{U})]\ge\mathsf{C}}I(Y;\tilde{U})$ is twice continuously differentiable as a function of $P_{Y}$ (let $P_{X,Y}=P_{X|Y}P_{Y}$, and $P_{U|Y}$ be the minimizer in $\mathrm{IB}(\mathsf{C})$), and perturbing $P_{Y}$ within a neighborhood of the original $P_{Y}$ will not affect the support of $U^{*}$, where $P_{U^{*}|Y}$ attains the minimum in $\tilde{R}(\mathsf{C})$.
}
We have
\begin{equation}
    \mathsf{B}_{\mathrm{V}}^{*}(n,\mathsf{C},\epsilon)\le(1-\epsilon)\!\left(\!\mathrm{IB}(\mathsf{C})+\sqrt{\frac{\ln n}{n}\mathrm{CVIB}(\mathsf{C})}\right)+O\!\left(\!\frac{1}{\sqrt{n}}\!\right)\!,
\end{equation}
where 
$\mathrm{IB}(\mathsf{C})=\mathrm{IB}_{X\to Y}(\mathsf{C})$ and $\mathrm{CVIB}(\mathsf{C})=\mathrm{CVIB}_{X\to Y}(\mathsf{C})$.
\end{thm}
\begin{IEEEproof}
Consider the $P_{U|Y}$ that achieves the minimum in $\mathrm{IB}(\mathsf{C})$ (for tie-breaking, choose the $P_{U|Y}$ that gives the smallest $\mathbb{E}[\mathrm{Var}[\iota_{X;U}(X,U)\,|\,Y,U]]$). Define a distortion function $d(x,u)=-\iota_{X;U}(x;u)$, and  consider the rate-distortion function $R(\mathsf{D})=\min_{P_{\tilde{U}|Y}:\,\mathbb{E}[d(X,\tilde{U})]\le\mathsf{D}}I(Y;\tilde{U})$ of the noisy lossy source coding problem 
at $\mathsf{D}=-\mathsf{C}$. Since
\ifshortver
\else
\begin{align}
& I(X;\tilde{U})-\mathbb{E}[\iota_{X;U}(X,\tilde{U})] \nonumber \\
& = \mathbb{E}\left[\log\frac{P_{X|\tilde{U}}(X|\tilde{U})}{P_{X|U}(X|\tilde{U})}\right] \nonumber \\
& = \mathbb{E}\left[D(P_{X|\tilde{U}}(\cdot|\tilde{U})\Vert P_{X|U}(\cdot|\tilde{U}))\right] \ge 0,\label{eq:pf_i_density_ineq}
\end{align}
\fi
$\mathbb{E}[d(X,\tilde{U})]\le\mathsf{D}$ implies $I(X;\tilde{U})\ge\mathsf{C}$, and hence $P_{U|Y}$ also achieves the minimum in $R(\mathsf{D})$, and $R(\mathsf{D})=\mathrm{IB}(\mathsf{C})$. \eqref{eq:pf_i_density_ineq} also implies that $R(\mathsf{D}-\delta)\ge\mathrm{IB}(\mathsf{C}+\delta)$ for every $\delta\in\mathbb{R}$, so we must have $-R'(\mathsf{D})=\mathrm{IB}'(\mathsf{C})=:\lambda^{*}$. Theorem \ref{thm:noisy_vl_2}  gives a noisy lossy coding scheme with
\begin{align}
& \mathbb{E}[|W|]  \nonumber\\
& \le(1-\epsilon)\Big(nI(Y;U) \nonumber\\
 & \;\;\;+\lambda^{*}\sqrt{(n\ln n)\mathbb{E}[\mathrm{Var}[d(X,U)\,|\,Y,U]]}\Big)+O(\sqrt{n}) \nonumber\\
 & \le(1-\epsilon)\Big(n\mathrm{IB}(\mathsf{C})+\sqrt{(n\ln n)\mathrm{CVIB}(\mathsf{C})}\Big)+O(\sqrt{n}),
\end{align}
with a decoded sequence $\hat{U}^{n}$ satisfying (see footnote \ref{fn:noisy_vl_2_pe}) 
\begin{equation}
\mathbb{P}\big(\iota_{X^{n};U^{n}}(X^{n};\hat{U}^{n})<n\mathsf{C}+\log n\big) \le \epsilon-1/\sqrt{n}.\label{eq:pf_lossy_pe2}
\end{equation}
We apply the \textcolor{black}{generalized Poisson matching lemma (Lemma~\ref{lem:general_pml}) \cite{li2021unified}} as in the proof of Theorem \ref{thm:ib_one_vl} on $X^{n},\hat{U}^{n}$ and $\mathsf{L}=\lceil2^{n\mathsf{C}}\rceil$, which for $n\ge 4$ gives
\ifshortver
\else
\begin{align}
& \mathbb{P}(M\neq\hat{M}|M=m)  \nonumber\\
& \le2^{-(n\mathsf{C}+\log n)}m+\epsilon-1/\sqrt{n}  \nonumber\\
 & \le2^{-(n\mathsf{C}+\log n)}(2^{n\mathsf{C}}+1)+\epsilon-1/\sqrt{n} \nonumber \\
 & \le2/n+\epsilon-1/\sqrt{n}\,\le\,\epsilon. 
\end{align} 
\fi
\end{IEEEproof}

\subsection{Fixed-length Description}

We now consider the case where $W\in[\textcolor{black}{\mathsf{L}_W}]$ is a fixed-length description. The following achievability result is a corollary of Theorem \ref{thm:noisy_fl} and the Poisson matching lemma \cite{li2021unified}. The proof is the same as that of Theorem \ref{thm:ib_one_vl} (except we use the fixed-length result in Theorem \ref{thm:noisy_fl} instead of the variable-length result in Theorem \ref{thm:noisy_vl}), and is omitted.

\smallskip{}
\begin{thm}
Fix $P_{X}$, $P_{U|Y}$ and $\mathsf{C},\gamma>0$. 
There is a one-shot fixed-length scheme with message size $\mathsf{L}$, description size $\textcolor{black}{\mathsf{L}_W}$,
\begin{equation}
    P_{e}\le\mathbb{P}\big(\psi_{U}(Y,\mathsf{C},T)\ge\log\gamma\big)+2^{-\mathsf{C}}(\mathsf{L}+1)/2+e^{-\textcolor{black}{\mathsf{L}_W}/\gamma},
\end{equation}
where $T\sim\mathrm{Unif}(0,1)$, $T\perp \!\!\!\perp Y$, with $\psi_{U}$ defined in \eqref{eq:psi_U}.
\end{thm}
\smallskip{}

We then give a second-order result in terms of the var-information bottleneck $\mathrm{VIB}(\mathsf{C})$ in \eqref{eq:VIB}.  \textcolor{black}{Refer to Appendix~\ref{subsec:pf_card} for a cardinality bound for the auxiliary random variable $U$ in the definition of $\mathrm{VIB}(\mathsf{C})$.}

\smallskip{}
\begin{thm} \label{thm:ib_n_fl}
Fix any $P_{X}$, $\epsilon>0$ and $0<\mathsf{C}<I(X;Y)$. Under the regularity conditions in Theorem \ref{thm:ib_n_vl}, we have
\begin{equation}
    \mathsf{B}_{\mathrm{F}}^{*}(n,\mathsf{C},\epsilon)\le\mathrm{IB}(\mathsf{C})+\sqrt{\frac{1}{n}\mathrm{VIB}(\mathsf{C})}Q^{-1}(\epsilon)+O\left(\frac{\log n}{n}\right),
\end{equation}
where 
$\mathrm{IB}(\mathsf{C})=\mathrm{IB}_{X\to Y}(\mathsf{C})$ and $\mathrm{VIB}(\mathsf{C})=\mathrm{VIB}_{X\to Y}(\mathsf{C})$.
\end{thm}
\begin{IEEEproof}
As in Theorem \ref{thm:ib_n_vl}, consider $P_{U|Y}$ that achieves the minimum in $\mathrm{IB}(\mathsf{C})$. 
Let $d(x,u)=-\iota_{X;U}(x;u)$ and $R(\mathsf{D})=\min_{P_{\tilde{U}|Y}:\,\mathbb{E}[d(X,\tilde{U})]\le\mathsf{D}}I(Y;\tilde{U})$. 
We have shown that $R(\mathsf{D})=\mathrm{IB}(\mathsf{C})$ and $-R'(\mathsf{D})=\mathrm{IB}'(\mathsf{C})=:\lambda^{*}$. Theorem \ref{thm:noisy_fl_2}  gives a noisy lossy coding scheme with decoded sequence $\hat{U}^{n}$ satisfying
\begin{align}
\log\textcolor{black}{\mathsf{L}_W} & \le n\mathrm{IB}(\mathsf{C})+\sqrt{n\mathrm{Var}[\iota_{Y;U}(Y;U)-\lambda^{*}\iota_{X;U}(X;U)]}  \nonumber\\
 & \;\;\;\;\;\cdot Q^{-1}(\epsilon)+O(\log n)  \nonumber\\
 & =n\mathrm{IB}(\mathsf{C})+\sqrt{n\mathrm{VIB}(\mathsf{C})}Q^{-1}(\epsilon)+O(\log n).
\end{align}
Inspecting \cite[Appendix D]{kostina2016nonasymptotic} shows that the bound $\mathbb{P}(d(X^{n},\hat{U}^{n})>\mathsf{D})\le\epsilon$ in Theorem \ref{thm:noisy_fl_2} can be strengthened to $\mathbb{P}(d(X^{n},\hat{U}^{n})>\mathsf{D}-n^{-1}\log n)\le\epsilon-1/\sqrt{n}$, giving the same bound as \eqref{eq:pf_lossy_pe2}. The proof is completed by applying the \textcolor{black}{generalized} Poisson matching lemma \cite{li2021unified} as in Theorem \ref{thm:ib_n_vl}.
\end{IEEEproof}

\section{Concluding Remarks}
We have shown nonasymptotic achievability results for the information bottleneck channel with fixed and variable-length descriptions, using techniques in noisy lossy source coding and Poisson functional representation. We have also shown novel bounds for variable-length noisy lossy source coding. For future directions, it is of interest to study converse results and investigate whether Theorems \ref{thm:noisy_vl_2}, \ref{thm:ib_n_vl} and \ref{thm:ib_n_fl} are tight.

\section{Acknowledgement}

This work was partially supported by two grants from the Research Grants Council of the Hong Kong Special Administrative Region, China [Project No.s: CUHK 24205621 (ECS), CUHK 14209823 (GRF)]. 
We would like to thank Prof. Ioannis Kontoyiannis for the insightful discussion. 
\textcolor{black}{
We would also like to thank the associate editor and the anonymous reviewers of this paper and the conference version~\cite{info_btl_isit} for their constructive
and helpful comments.
}



\ifshortver
\else

\appendix{}

\subsection{Cardinality Bounds for $\mathrm{VIB}$ and $\mathrm{CVIB}$\label{subsec:pf_card}}
{\color{black}
As defined in \eqref{eq:VIB}, the var-information bottleneck is 
$\mathrm{VIB}(\mathsf{C})= \min_{P_{U|Y}} \mathrm{Var}[\iota_{Y;U}(Y;U)-\lambda^{*}\iota_{X;U}(X;U)]$, where $P_{U|Y}$ ranges over minimizers of $I(Y;U)$ under the constraint $I(X;U)\geq \mathsf{C}$ when $X\to Y\to U$. We now show that it suffices to consider $P_{U|Y}$ with the cardinality bound $|\mathcal{U}| \leq |\mathcal{Y}|+2$ using the standard approach via the Fenchel-Eggleston-Carath\'{e}odory (FEC) theorem \cite{elgamal2011network}. For $u \in \mathcal{U}$, consider the vector $\mathbf{v}_u \in \mathbb{R}^{|\mathcal{Y}|+2}$ consisting of entries $H(Y|U=u)$, $H(X|U=u)$, 
\begin{align}
\mathbb{E}[(\iota_{Y;U}(Y;U)-\lambda^{*}\iota_{X;U}(X;U))^2\,|\,U=u],\label{eq:pf_E_square}
\end{align}
and $P_{Y|U}(y|u)$ for $y \in \mathcal{Y} \backslash \{y_0\}$ (where $y_0\in \mathcal{Y}$ is arbitrarily chosen). By the FEC theorem, there is a distribution $P_{\tilde{U}}$ over $\mathcal{U}$ with cardinality $|\tilde{\mathcal{U}}| \leq |\mathcal{Y}|+2$ such that $\mathbb{E}[\mathbf{v}_U] = \mathbb{E}[\mathbf{v}_{\tilde{U}}]$.
If we generate $(\tilde{U},\tilde{Y}, \tilde{X})   \sim P_{\tilde{U}} P_{Y|U} P_{X|Y}$, then $P_{\tilde{Y}} = P_Y$, and hence $P_{\tilde{Y}, \tilde{X}} = P_{Y,X}$. Since $\mathbb{E}_{u \sim P_U}[H(Y|U=u)] = \mathbb{E}_{\tilde{u} \sim P_{\tilde{U}}}[H(Y|U=\tilde{u})]$, we have $H(\tilde{Y}|\tilde{U})=H(Y|U)$ and $I(\tilde{Y};\tilde{U})=I(Y;U)$. Similarly, $I(\tilde{X};\tilde{U})=I(X;U)$. Also, we have $\iota_{\tilde{Y};\tilde{U}}(y;u) = \iota_{\tilde{Y}}(y)-\iota_{\tilde{Y}|\tilde{U}}(y|u)=\iota_{Y}(y)-\iota_{Y|U}(y|u)=\iota_{Y;U}(y;u)$, and hence $\mathbb{E}[(\iota_{\tilde{Y};\tilde{U}}(\tilde{Y};\tilde{U})-\lambda^{*}\iota_{\tilde{X};\tilde{U}}(\tilde{X};\tilde{U}))^2] = \mathbb{E}[(\iota_{Y;U}(Y;U)-\lambda^{*}\iota_{X;U}(X;U))^2]$ by \eqref{eq:pf_E_square}. Therefore, $\mathrm{Var}[\iota_{\tilde{Y};\tilde{U}}(\tilde{Y};\tilde{U})-\lambda^{*}\iota_{\tilde{X};\tilde{U}}(\tilde{X};\tilde{U})] = \mathrm{Var}[\iota_{Y;U}(Y;U)-\lambda^{*}\iota_{X;U}(X;U)]$.

As defined in \eqref{eq:CVIB}, the conditional var-information bottleneck is 
$\mathrm{CVIB}(\mathsf{C})= \min_{P_{U|Y}} \mathbb{E}[\mathrm{Var}[\lambda^{*}\iota_{X;U}(X;U)\,|\,Y,U]]$, where the range of $P_{U|Y}$ is the same as in $\mathrm{VIB}(\mathsf{C})$.
The proof of the cardinality bound $|\mathcal{Y}|+2$ is the same as that for $\mathrm{VIB}(\mathsf{C})$, except that we use $\mathbb{E}[\mathrm{Var}[\lambda^{*}\iota_{X;U}(X;U)\,|\,Y,U] | U=u]$ instead of \eqref{eq:pf_E_square}.

}

\subsection{Remainder of the Proof of Theorem \ref{thm:noisy_vl}\label{subsec:pf_noisy_vl}}

We use the strategy in \cite[Theorem 2]{sfrl_trans} to remove the common randomness $G:=(\bar{Z}_{i},T_{i})_{i}$. 
\textcolor{black}{Although there are infinitely many values of $G$, since we are only interested in two quantities $\mathbb{E}[|W|]$ and $\mathbb{P}(d(X,Z)>\mathsf{D})$, using Carath\'{e}odory's theorem, we can find a discrete distribution over only two values of $G$ which preserves or reduces $\mathbb{E}[|W|]$ and $\mathbb{P}(d(X,Z)>\mathsf{D})$. More precisely, we can find}
two values $g_{0},g_{1}$ of the common randomness and $\lambda_{0}\in[0,1]$ (let $\lambda_{1}=1-\lambda_{0}$) such that 
\begin{align}
    & \sum_{i=0}^{1} \lambda_{i}\mathbb{E}\big[|W|\,\big|\,G=g_{i}\big]\le\mathbb{E}[|W|],\\
    & \sum_{i=0}^{1} \lambda_{i}\mathbb{E}\big[|W|\,\big|\,G=g_{i}\big]\le\mathbb{E}[|W|],
\end{align}
\begin{equation}
    \sum_{i=0}^{1} \lambda_{i}\mathbb{P}\big(d(X,Z)>\mathsf{D}\,\big|\,G=g_{i}\big)\le\mathbb{P}(d(X,Z)>\mathsf{D}).
\end{equation}
This is possible by Carath\'{e}odory's theorem \textcolor{black}{(see \cite[Theorem 2]{sfrl_trans})}. The encoder generates $J\sim\mathrm{Bern}(\lambda_{1})$ and transmits it to the decoder, and then \textcolor{black}{performs} the aforementioned encoding scheme conditional on $G=g_{J}$. The decoder receives $J,W$ and \textcolor{black}{performs} the decoding scheme conditional on $G=g_{J}$. Since $J$ takes one bit to transmit, the resultant expected length is 
\begin{align}
    & \mathbb{E}[|W|]+1 \nonumber \\
    & \le H(\tilde{K})+2 \nonumber \\
    & \le\ell(\mathbb{E}[(1-\beta(Y))\psi_{\bar{Z}}(y,\mathsf{D},\epsilon')]).
\end{align}

\subsection{Proof of Theorem \ref{thm:noisy_vl_2}\label{subsec:pf_noisy_vl_2}}

Define $\phi(y^{n},z^{n},\mathsf{D}):=\mathbb{P}(d(X^{n},z^{n})>\mathsf{D}|Y^{n}=y^{n})$. Fix $\epsilon_{1}>0$. It was shown in \cite[Appendix D]{kostina2016nonasymptotic} that  for all typical $y^{n}\in\mathcal{T}_{n}$ (where $\mathcal{T}_{n}:=\{y^{n}:\,\Vert\hat{P}_{y^{n}}-P_{Y}\Vert^{2}\le|\mathcal{Y}|n^{-1}\log n\}$, and $\hat{P}_{y^{n}}$ is the empirical distribution of $y^{n}$),
\begin{align}
 & \psi_{Z^{*n}}(y^{n},\mathsf{D},\epsilon_{1})\nonumber \\
 & =\inf_{P_{Z^{n}}:\,\phi(y^{n},Z^{n},\mathsf{D})\le\epsilon_{1}\;\mathrm{a.s.}}D(P_{Z^{n}}\Vert P_{Z^{*}}^{n})\nonumber \\
 & \le\sum_{i=1}^{n}\jmath(y_{i},\mathsf{D})+\lambda^{*}\sqrt{nV_{d}}Q^{-1}(\epsilon_{1})+O(\log n),\label{eq:pf_psi1}
\end{align}
where $\jmath(y,\mathsf{D}):=\iota_{Y;Z^{*}}(y;z)+\lambda^{*}(\mathbb{E}[d(X,z)|Y=y]-\mathsf{D})$ (this holds for $P_{Z^{*}}$-almost all $z$) is the $d$-tilted information for the corresponding noiseless source coding problem, and $V_{d}:=\mathbb{E}[\mathrm{Var}[d(X,Z^{*})\,|\,Y,Z^{*}]]$. For footnote \ref{fn:noisy_vl_2_pe}, note that the arguments in \cite[Appendix D]{kostina2016nonasymptotic} show that \eqref{eq:pf_psi1} continues to hold if $\psi_{Z^{*n}}(y^{n},\mathsf{D},\epsilon_{1})$ in the left hand side is replaced with $\psi_{Z^{*n}}(y^{n},\mathsf{D}-n^{-1}\log n,\epsilon_{1})$, which does not affect the use of the Berry-Esse\'{e}n theorem. We have \cite{kostina2016nonasymptotic}
\ifshortver
\begin{equation}
\mathbb{P}(Y^{n}\notin\mathcal{T}_{n})\le2|\mathcal{Y}|/\sqrt{n}=:\epsilon_{2}-1/\sqrt{n},\label{eq:pf_typical}
\end{equation}
\else
\begin{align}
& \mathbb{P}(Y^{n}\notin\mathcal{T}_{n}) \nonumber\\
& \le 2|\mathcal{Y}|/\sqrt{n}=:\epsilon_{2}-1/\sqrt{n}, 
\end{align}
\fi
where $\epsilon_{2}:=(2|\mathcal{Y}|+1)/\sqrt{n}$. Take $\beta(y^{n})=1$ if $y^{n}\notin\mathcal{T}_{n}$, and $\beta(y^{n})=\epsilon_{3}$ if $y^{n}\in\mathcal{T}_{n}$. By \eqref{eq:pf_psi1}, letting $\epsilon=\epsilon_{1}+\epsilon_{2}+\epsilon_{3}$ and $R:=R(\mathsf{D})=\mathbb{E}[\jmath(Y,\mathsf{D})]$,
\begin{align}
 & \mathbb{E}[\psi_{Z^{*n}}(Y^{n},\epsilon_{1})(1-\beta(Y^{n})]\nonumber \\
 & \le(1\!-\!\epsilon_{3})\!\left(\!\mathbb{E}\!\left[\sum_{i=1}^{n}\jmath(Y_{i},\mathsf{D})\right]+\lambda^{*}\sqrt{nV_{d}}Q^{-1}(\epsilon_{1})\!\!\right)\!+O(\log n)\nonumber \\
 & =(1-\epsilon_{3})\left(nR+\lambda^{*}\sqrt{nV_{d}}Q^{-1}(\epsilon_{1})\right)+O(\log n).\label{eq:pf_psi_bd}
\end{align}
Take $\epsilon_{1}=1/(2\sqrt{n})$. We have
\begin{align}
(1-\epsilon_{3})nR & =(1-\epsilon)nR+\epsilon_{1}nR+\epsilon_{2}nR\nonumber \\
 & =(1-\epsilon)nR+(R/2)\sqrt{n}+(2|\mathcal{Y}|+1)R\sqrt{n}\nonumber \\
 & =(1-\epsilon)nR+O(\sqrt{n}).\label{eq:pf_nr_bd}
\end{align}
By Chernoff bound $Q(t)\le e^{-t^{2}/2}$, we have
\begin{align}
\lambda^{*}\sqrt{nV_{d}}Q^{-1}(\epsilon_{1}) & \le\lambda^{*}\sqrt{2nV_{d}\ln(1/\epsilon_{1})} \nonumber\\
 & =\lambda^{*}\sqrt{V_{d}n\ln n}+O\left(\sqrt{\frac{n}{\ln n}}\right),
\end{align}
and
\begin{align}
 & (1-\epsilon_{3})\lambda^{*}\sqrt{nV_{d}}Q^{-1}(\epsilon_{1})\nonumber \\
 & \le(1-\epsilon_{3})\lambda^{*}\sqrt{V_{d}n\ln n}+O\left(\sqrt{\frac{n}{\ln n}}\right)\nonumber \\
 & =(1-\epsilon)\lambda^{*}\sqrt{V_{d}n\ln n}+O\left(\sqrt{\frac{n}{\ln n}}\right).\label{eq:pf_dist_bd}
\end{align}
Substituting \eqref{eq:pf_nr_bd} and \eqref{eq:pf_dist_bd} into \eqref{eq:pf_psi_bd},
\begin{align}
 & \mathbb{E}[\psi_{Z^{*n}}(Y^{n},\mathsf{D},\epsilon_{1})(1-\beta(Y^{n})] \nonumber\\
 & \le(1-\epsilon)\left(nR+\lambda^{*}\sqrt{V_{d}n\ln n}\right)+O(\sqrt{n}).
\end{align}
Invoking Theorem \ref{thm:noisy_vl}, there exists a variable-length code with
\begin{align}
\mathbb{E}[|W|] & \le\ell\left(\mathbb{E}[\psi_{Z^{*n}}(Y^{n},\mathsf{D},\epsilon_{1})(1-\beta(Y^{n})]\right) \nonumber\\
 & \le(1-\epsilon)\left(nR+\lambda^{*}\sqrt{V_{d}n\ln n}\right)+O(\sqrt{n})+O(\log n) \nonumber\\
 & =(1-\epsilon)\left(nR+\lambda^{*}\sqrt{V_{d}n\ln n}\right)+O(\sqrt{n}),
\end{align}
and
\begin{align}
P_{e} & \le\mathbb{E}[\beta(Y^{n})]+\epsilon_{1} \nonumber\\
 & \le\mathbb{P}(Y^{n}\notin\mathcal{T}_{n})+\epsilon_{3}+\epsilon_{1} \nonumber\\
 & \le\epsilon_{2}-1/\sqrt{n}+\epsilon_{3}+\epsilon_{1} \nonumber\\
 & = \epsilon-1/\sqrt{n}.
\end{align}

\fi

\bibliographystyle{IEEEtran}
\bibliography{ref}

\end{document}